\documentclass[11pt]{article}
\usepackage{amssymb}
\usepackage{amsmath}
\usepackage{amscd}
\usepackage{latexsym}

\catcode`@=11
\renewcommand{\section}{\@startsection{section}{1}{0pt}{\medskipamount}
{\medskipamount}{\large\bf}}
\numberwithin{equation}{section}
\catcode`@=12

\def\a{\alpha}
\def\b{\beta}
\def\D{\Delta}
\def\de{\delta}

\def\ve{\varepsilon}
\def\h{\eta}
\def\th{\theta}

\def\la{\lambda}
\def\m{\mu}
\def\n{\nu}
\def\r{\rho}
\def\s{\sigma}

\def\vp{\varphi}

\def\La{\Lambda}

\newcommand{\hi}{{\hat\imath}}
\newcommand{\hj}{{\hat\jmath}}
\newcommand{\unity}{{\mathbf{1\hspace{-2.9pt}l}}}
\newcommand{\zero}{{\mathbf{0}}}
\newcommand{\bx}{{\mathbf{x}}}
\newcommand{\ba}{{\mathbf{a}}}
\newcommand{\bb}{{\mathbf{b}}}
\newcommand{\bff}{{\mathbf{f}}}
\def\mi{{\mu_\hi}}
\def\ni{{\nu_\hi}}

\def\nj{{\nu_\hj}}

\newcommand{\C}{\mathbb C}
\newcommand{\R}{\mathbb R}
\newcommand{\Z}{\mathbb Z}
\newcommand{\Hbb}{{\mathbb H}}
\newcommand{\Hcal}{{\cal H}}
\newcommand{\HP}{{\mathbb H}{P}}

\def\ic{\mbox{i}}
\def\N2{$N{=}2$}
\def\pa{\mbox{$\partial$}}
\def\diff{\mbox{d}}
\def\tr{{\rm tr}}
\def\sfrac#1#2{{\textstyle\frac{#1}{#2}}}
\def\>{\rangle}
\def\<{\langle}
\def\+{\dagger}
\def\={\ =\ }
\def\und{\qquad\textrm{and}\qquad}

\begin{document}

\begin{titlepage}
\setcounter{page}{0}
\begin{flushright}
hep-th/0703009\\
ITP--UH--03/07\\
\end{flushright}

\vskip 2.0cm

\begin{center}

{\Large\bf Construction of Noncommutative Instantons in 4k Dimensions}

\vspace{15mm}

{\Large Johannes Br\"odel${}^*$, 
Tatiana A. Ivanova${}^\+$ \ and \ 
Olaf Lechtenfeld${}^*$ }
\\[10mm]
\noindent ${}^*${\em Institut f\"ur Theoretische Physik,
Leibniz Universit\"at Hannover \\
Appelstra\ss{}e 2, 30167 Hannover, Germany }\\
{Email: jbroedel, lechtenf@itp.uni-hannover.de}
\\[10mm]
\noindent ${}^\+${\em Bogoliubov Laboratory of Theoretical Physics, JINR\\
141980 Dubna, Moscow Region, Russia}\\
{Email: ita@theor.jinr.ru}
\vspace{20mm}

\begin{abstract}
\noindent
We consider generalized self-duality equations for U($2r$) Yang-Mills theory 
on $\R^{4k}$ with quaternionic structure and self-dual Moyal deformation. 
We employ the extended ADHM method in $4k$ dimensions to construct new 
noncommutative generalizations of the 't~Hooft as well as of the BPST 
instantons. It is shown that in the commutative limit the BPST-type 
configurations coincide with the standard instantons on ${\HP}^n$ written 
in local coordinates.

\end{abstract}

\end{center}
\end{titlepage}

\section{Introduction}

\noindent
Moyal-type deformations of gauge field theories arise naturally in string 
theory in the presence of D-branes and a $B$-field background~\cite{Sei}. 
Such theories are also interesting by themselves as specific nonlocal 
generalizations of ordinary gauge theories. Among classical solutions 
of the Moyal-deformed Yang-Mills theory in four dimensions, of particular
importance are noncommutative instantons (see e.g.~\cite{gr1, LP, gr2} and 
references therein), which are BPS configurations solving the self-dual 
Yang-Mills (SDYM) equations~\cite{BPST, Ward}.

Natural BPS-type equations in more than four dimensions~\cite{CW, DU}, 
known as generalized self-duality or generalized SDYM equations, appear in 
superstring compactification as the conditions for the survival of at least 
one supersymmetry~\cite{GSW88}. Various solutions to these first-order 
equations were found e.g.~in~\cite{FNP, GFKL}, and their noncommutative 
generalizations have been considered e.g.~in~\cite{ncgen, Po, IL05}. Some 
of these works employ the ADHM solution technique~\cite{ADHM} which was 
generalized from four-dimensional to $4k$-dimensional quaternionic spaces 
in~\cite{CGK}. 

In this letter we consider Yang-Mills theory on noncommutative 
$\Hbb^k\cong\R^{4k}$, where $\Hbb$ is the algebra of quaternions. 
Up to Sp($k$) rotations, a self-dual Moyal deformation tensor in $4k$ 
dimensions is characterized by real parameters $\th_0,\th_1,\dots,\th_{k-1}$,
one for each four-dimensional subspace.
Using the extended ADHM construction, we find two kinds of explicit solutions, 
which generalize the 't~Hooft and BPST instantons from $\R^4$ to self-dually 
noncommutative $\R^{4k}$. 
In the first ('t~Hooft-type) case, our solutions univeralize the ones obtained 
in~\cite{IL05} for a special choice of $(\th_0,\th_1,\dots,\th_{k-1})$ to an
arbitrary choice. Note that four-dimensional ``slices'' of these solutions 
coincide with the noncommutative $n$-instanton configurations obtained 
in~\cite{LP} or~\cite{ILM} depending on the value of the 
noncommutativity parameters.  
In the second (BPST-type) case, we generalize the solution obtained 
in~\cite{IL05} from $\R^8_\th$ to $\R^{4k}_\th$ for arbitrary $k$. 
In particular, for $k{=}1$ this solution coincides with the noncommutative
deformation~\cite{gr2} of the BPST instanton~\cite{BPST}.

\bigskip

\section{Generalized self-duality on $\Hbb^{k}$}

\noindent
Let us consider the $4k$-dimensional space $\R^{4k}$ with the metric 
$\de_{\m\n}$, where $\m,\n,\ldots=1,\ldots,4k$. It
can be decomposed into a direct sum of $k$ four-dimensional spaces,
\begin{equation}\label{R4k}
\R^{4k}\ \cong\ \R^{k}\otimes\R^{4}\=\R^{4}\oplus\ldots\oplus\R^{4}\=
\Hbb\oplus\ldots\oplus\Hbb\=\Hbb^k\ ,
\end{equation}
with coordinates $x^\mi$, where 
$\mi= 4\hi{+}\m_0 \in\{4\hi{+}1,\ldots,4\hi{+}4\}$ 
is a four-valued index and $\hi=0,\ldots,k{-}1$.
Each such subspace is identified with the algebra $\Hbb$
of quaternions realized in terms of $2{\times}2$ matrices with a basis
(same for all values of $\hi$)
\begin{equation}\label{emu}
\bigl(e_\mi^\+\bigr) \=
\bigl(\ic\,\s_1\,, \ic\,\s_2\,, \ic\,\s_3\,,\unity_2\bigr)
\quad\Rightarrow\quad
\bigl(e_\mi\bigr) \=
\bigl(-\ic\,\s_\a\,,\unity_2\bigr)\ ,
\end{equation}
where $\s_\a$, $\a{=}1,2,3$, are the Pauli matrices. We define a basis 
$V_\m$ with $(\m)=(\m_0,\ldots,\m_{k-1})$ on $\Hbb^k$ as a collection of $k$
quaternionic column vectors realized as $2k{\times}2$ matrices
\begin{equation}\label{Vmu}
V_{\m_0} = \begin{pmatrix}
e^\+_{\m_0}\\ \zero_2 \\ \vdots \\ \vdots \\ \zero_2 \end{pmatrix}
\ ,\quad\ldots\ ,\quad
V_{\mi} = \begin{pmatrix}
\zero_2 \\ \vdots \\ e^\+_\mi \\ \vdots \\ \zero_2 \end{pmatrix}
\ ,\quad\ldots\ ,\quad
V_{\m_{k-1}} = \begin{pmatrix}
\zero_2 \\ \vdots \\ \vdots \\ \zero_2 \\ e^\+_{\m_{k-1}} \end{pmatrix}
\end{equation}
with $e^\+_\mi$ in the $\hi$-th position. A point 
$(x^\m)\in\R^{4k}$ can be represented by the $2k{\times}2$ matrix
\begin{equation}\label{bfx}
{\bx}\=x^\m V_\m\=x^{\m_0} V_{\m_0}+\ldots +x^{\mu_{k-1}}V_{\mu_{k-1}}\=
\begin{pmatrix}x^{\m_0}e^\+_{\m_0}\\ \vdots\\ x^{\m_{k-1}}e^\+_{\m_{k-1}}
\end{pmatrix}\ =:\ \begin{pmatrix}x_0\\ \vdots\\ x_{k-1}\end{pmatrix}\ .
\end{equation}

Note that the matrices (\ref{emu}) have the properties
\begin{equation}
\label{prpts}
\begin{aligned}
e^\+_\mi e_\nj \= e^\+_{\m_0} e_{\n_0} \ =&\
\de_{\m_0\n_0}\ \unity_2\ +\ \eta^{\a}_{\m_0\n_0}\ \ic\,\s_\a\ =:\ 
\de_{\m_0\n_0}\ \unity_2\ +\ \eta_{\m_0\n_0}\ , \\
e_\mi e^\+_\nj \= e_{\m_0} e^\+_{\n_0} \ =&\
\de_{\m_0\n_0}\ \unity_2\ +\ \bar\eta^{\a}_{\m_0\n_0}\ \ic\,\s_\a\ =:\ 
\de_{\m_0\n_0}\ \unity_2\ +\ \bar\eta_{\m_0\n_0}\ ,
\end{aligned}
\end{equation}
where $\eta^{\a}_{\m_0\n_0}$ and $\bar\eta^{\a}_{\m_0\n_0}$ 
denote the self-dual and anti-self-dual 't~Hooft tensors, respectively, 
with $\m_0$ and $\n_0$ running from 1 to 4. Recall that the matrices
$\eta_{\m_0\n_0}$ and $\bar\eta_{\m_0\n_0}$ play an important role in the 
ADHM method~\cite{ADHM} of constructing self-dual and anti-self-dual solutions
of the Yang-Mills equations in four dimensions. Analogs of these matrices 
in $4k$ dimensions can be defined with the help of the matrices~(\ref{Vmu}).
Specifically, 
we introduce antihermitian $2k{\times}2k$ and $2{\times}2$ matrices~\cite{CGK}
\begin{equation}\label{N}
N_{\m\n}\ :=\ \sfrac12(V_\m V_\n^\+ - V_\n V_\m^\+) \und
\bar N_{\m\n}\ :=\ \sfrac12(V_\m^\+ V_\n - V_\n^\+ V_\m)\ ,
\end{equation}
which for $k{=}1$ coincide with the $2{\times}2$ matrices 
$\eta_{\m\n}$ and $\bar\eta_{\m\n}$, respectively. 
Notice that for any $\m,\n=1,\ldots,4k$ we have
$N_{\m\n}\in sp(k)\subset u(2k)$ and $\bar N_{\m\n}\in sp(1)\subset u(2)$.
For $k>1$ there also  exists a third tensor $M_{\mu\nu}$
taking values in the complement of $sp(1)\oplus sp(k)$ in $so(4k)$~\cite{CGK}.

To introduce generalized self-duality equations on $\R^{4k}$~\cite{CW, CGK} 
we define a tensor 
\begin{equation}\label{Q}
Q_{\m\n\r\s}\ :=\ \tr\,(V_\m^\+ V_\n V_\r^\+ V_\s)
\end{equation}
and its total antisymmetrization
\begin{equation}\label{T}
T_{\m\n\r\s}\ :=\ \sfrac{1}{12}Q_{\m[\n\r\s ]}\= \sfrac{1}{12}
(Q_{\m\n\r\s} + Q_{\m\s\n\r} + Q_{\m\r\s\n}
-Q_{\m\r\n\s} - Q_{\m\n\s\r} - Q_{\m\s\r\n})\ .
\end{equation}
Then by direct calculations one finds~\cite{CGK} that the matrix-valued 
tensor~$N_{\m\n}$ is self-dual in a generalized sense, 
i.e. it satisfies the eigenvalue equations
\begin{equation}\label{TN}
\sfrac12 T_{\m\n\r\s} N_{\r\s} \= N_{\m\n} \ , \qquad\textrm{while}\qquad
\sfrac12 T_{\m\n\r\s} \bar N_{\r\s} \= -\sfrac{2k+1}{3} \bar N_{\m\n}
\end{equation}
shows that $\bar N_{\m\n}$ is anti-self-dual only for $k{=}1$.
It is well known that the subgroup of SO($4k$) which preserves the quaternionic 
structure and therefore (\ref{TN}) is isomorphic to Sp(1)$\times$Sp($k$)$/\Z_2$.

With the help of the tensor (\ref{T}) one may introduce an analog of the SDYM 
equations for U($2r$) gauge fields on $\R^{4k}$. 
Namely, we consider a $u(2r)$-valued gauge potential $A=A_\m\diff x^\m$ 
and the Yang-Mills field $F_{\m\n}=\pa_\m A_\n -\pa_\n A_\m + [A_\m ,A_\n ]$,  
where $(x^\m)\in\R^{4k}$ and $\pa_\m :=\pa /\pa x^\m$. 
The generalized SDYM equations on $\R^{4k}$ read~\cite{CW,CGK}
\begin{equation}\label{gsdym}
\sfrac{1}{2}T_{\mu\nu\r\s} F_{\r\s} \= F_{\mu\n}
\end{equation}
and are clearly invariant under Sp(1)$\times$Sp($k$)$/\Z_2\subset$ SO($4k$).
Obviously, any gauge field fulfilling (\ref{gsdym}) 
satisfies the second-order Yang-Mills equations due to the Bianchi identities.
We remark that (\ref{gsdym}) can be derived as compatibility conditions of a 
linear system of equations, and its solutions can be constructed via a 
(generalized) twistor approach~\cite{CW,CGK}. 
In four dimensions $T_{\m\n\r\s}$ reduces to $\ve_{\m\n\r\s}$
and, hence, (\ref{gsdym}) coincides with the standard SDYM equations.

\bigskip

\section{Extended ADHM construction in 4k dimensions}

\noindent
The most systematic way to generate solutions to the generalized SDYM equations
(\ref{gsdym}) is via the ADHM approach extended to $4k$ dimension~\cite{CGK}. 
For $u(2r)$-valued gauge fields this method is based on
\begin{equation} \label{Psi}
\textrm{a complex }\quad(2\ell{+}2r)\times 2r\quad \textrm{matrix}\quad \Psi
\qquad\textrm{and}
\end{equation}
\begin{equation}\label{Delta}
\textrm{a complex }\quad(2\ell{+}2r)\times2\ell \quad\textrm{matrix }\quad
\Delta\={\ba}+{\bb}\ ({\bx} {\otimes} \unity_\ell)\=
{\ba}\ +\ \sum\limits^{k-1}_{\hi=0} {\bb}_\hi\,(x_\hi {\otimes} \unity_\ell)\ ,
\end{equation}
where $\ba$ and~${\bb}_\hi$ are constant $(2\ell{+}2r)\times 2\ell$ matrices
containing the size and position moduli.\footnote{
Note that $\bb$ is a constant $(2\ell{+}2r)\times2k\ell$ matrix,
$\bx$ is the $2k\times 2$ matrix given in~(\ref{bfx}) and
$x_\hi=x^\mi e^\+_\mi$ is $2\times 2$.
Correspondingly, ${\bx}\otimes{\unity}_\ell$ and $x_\hi\otimes{\unity}_\ell$ 
are $2k\ell\times 2\ell$ and $2\ell\times 2\ell$ matrices, respectively.}
These matrices must satisfy the following conditions:
\begin{eqnarray}
\Delta^\+\Delta\ &=&\ \bff^{-1}
\qquad\quad\ \textrm{(invertibility)}\ ,\label{c1}\\
{}[\,\Delta^\+\Delta\,,\,V_\m\otimes{\unity}_\ell\,]\ &=&\ 0 \ ,\label{c2}\\
\Delta^\+\Psi\ &=&\ 0\ ,\label{c3}\\
\Psi^\+\Psi\ &=&\ {\unity}_{2r}
\qquad\quad\,\,\textrm{(normalization)}\ ,\label{c4}\\
\Psi\,\Psi^\+\ +\ \Delta\,\bff\,\Delta^\+\ &=&\ {\unity}_{2\ell+2r}
\qquad\textrm{(completeness)}\ .\label{complete}
\end{eqnarray}
The completeness relation (\ref{complete}) means that $\Psi\,\Psi^\+$ and 
$\Delta\,\bff\,\Delta^\+$ are projectors onto orthogonal
complementing subspaces of $\C^{2\ell +2r}$.

Given a pair $(\Delta,\Psi)$ satisfying~(\ref{c1})--(\ref{complete}),
a (generalized) self-dual gauge potential arises from
\begin{equation} \label{adhmA}
A\= \Psi^\+\,\diff\Psi\qquad\Rightarrow\qquad A_\m\= \Psi^\+\,\pa_\m\Psi\ .
\end{equation}
Indeed, after straightforward calculation the components of the gauge field~$F$
then take the form
\begin{equation} 
\label{adhmF}
F_{\mu\nu}\=
\pa_\m(\Psi^\+\pa_\n\Psi)\ -\ \pa_\n(\Psi^\+\pa_\m\Psi)\
+\ [\,\Psi^\+\pa_\m\Psi\,,\,\Psi^\+\pa_\n\Psi\,]\ =\ 
2\,\Psi^\+ {\bb}\,N_{\m\n}\,\bff\,{\bb}^\+\Psi
\end{equation}
which, due to (\ref{TN}), 
is seen to satisfy the generalized SDYM equations (\ref{gsdym}).

\bigskip

\section{Self-dual noncommutative deformation}

\noindent
Classical field theory on the noncommutative deformation $\R^{4k}_\th$ of the 
space $\R^{4k}$ may be realized in a star-product formulation or in an operator 
formalism~\cite{Sei}. The nonlocality of the star product renders explicit 
computations cumbersome. We therefore take advantage of the Moyal-Weyl
correspondence and pass to the operator formalism, which trades the star
product for operator-valued coordinates $\hat x^\mu$ satisfying the
Heisenberg-algebra relations
\begin{equation} \label{nccoord}
[\,\hat x^\mu\,,\,\hat x^\nu\,] \= \ic\,\th^{\mu\nu} 
\end{equation}
with a constant real antisymmetric tensor $\th^{\m\n}$. 
The operators $\hat x^\m$ (and thus all their functions) act in an auxiliary 
Fock space~$\Hcal = \Hcal_0\otimes\Hcal_1\otimes\ldots\otimes\Hcal_{k-1}$, 
where each $\Hcal_\hi$ is a two-oscillator Fock space.  

The extended ADHM construction described in the previous section carries over 
to the noncommutative realm without change only for a self-dual deformation, 
i.e. if
\begin{equation}\label{4.2}
\sfrac12 T_{\m\n\r\s} \th_{\r\s} \= \th_{\m\n} 
\qquad\Leftrightarrow\qquad 
\th_{\m\n} \= \tr\,(N_{\m\n}\th)
\qquad\textrm{for}\quad \th_{\m\la}\th^{\la\n}=\de_\m^\n\ ,
\end{equation}
where $\th$ is an arbitrary $sp(k)$ matrix of parameters. We may employ an
Sp$(k)\subset$~SO$(4k)$ basis transformation to bring $\th$ to the diagonal form
\begin{equation}\label{4.3}
\th\=-\sfrac{\ic}2\,\mbox{diag}(\th_0\s_3,\,\th_1\s_3,\ldots,\,\th_{k-1}\s_3)\ ,
\end{equation}
which is characterized by real numbers $(\th_0,\th_1,\ldots,\th_{k-1})$.
{}From (\ref{4.2}) and (\ref{4.3}) we obtain
\begin{equation} \label{comrel}
\th^{\mi\ni}\=\th_\hi\,\h^{3\m_0\n_0} \qquad\textrm{and zero otherwise.}
\end{equation}
We drop the hats over operators from now on and assume that
all $\th_{\hat\imath}\ge 0$; 
the general case does not hide additional complication.

We introduce an operator version $(y_\hi, z_\hi)$ of the complex coordinates 
on $\R^{4k}\cong\C^{2k}$ via the formulae
\begin{equation}\label{4.5}
x_\hi \= x^\mi\, e^\+_\mi \ =:\ \ic 
\begin{pmatrix}\bar z_\hi & \phantom{-}\bar y_\hi\\ y_\hi & -z_\hi\end{pmatrix}
\qquad\Rightarrow\qquad
x_\hi^\+ \= -\ic
\begin{pmatrix}z_\hi & \phantom{-}\bar y_\hi\\ y_\hi & -\bar z_\hi\end{pmatrix}
\end{equation}
which imply
\begin{equation}\label{4.6}
[y_\hi\, ,\ \bar y_\hj] \= 2\,\th_\hi\,\de_{\hi\hj} \und
[z_\hi\, ,\ \bar z_\hj] \= 2\,\th_\hi\,\de_{\hi\hj}
\end{equation}
and the vanishing of all other commutators. 
Thus, the two-oscillator Fock space $\Hcal_{\hat\imath}$
is spanned by the basis states
\begin{equation}\label{4.7}
|m,n\>_{\hat\imath} \= 
\left( (2\th_\hi)^{m+n}m!n!\right)^{-1/2} \bar y_\hi^m \bar z_\hi^n\ |0,0\>_\hi
\qquad\textrm{for}\quad m,n=0,1,2,\ldots\ , 
\end{equation}
and the basis states in the full Fock space 
$\Hcal=\mathop{\otimes}\limits^{k-1}_{\hi=0}\Hcal_\hi$ take the form
\begin{equation}\label{4.8}
|m_0, n_0; m_1, n_1; \ldots ; m_{k-1}, n_{k-1}\> \= 
\mathop{\otimes}\limits^{k-1}_{\hi=0} |m_\hi,n_\hi\>_\hi\ . 
\end{equation}
Although the operator-valued coordinates (\ref{4.5}) are defined to act on
$\Hcal_\hi$ only, it is understood that their action extends to the full
Fock space $\Hcal$ via \
$x_\hi\ \to\ \unity_2\otimes\cdots\otimes x_\hi\otimes\cdots\unity_2$.
Note that if some $\th_\hi\le0$ then one should simply interchange the 
corresponding creation and annihilation operators, i.e.
$\bar y_\hi\leftrightarrow y_\hi$ and $\bar z_\hi\leftrightarrow z_\hi$.

\bigskip

\section{Noncommutative 't~Hooft-type solutions in 4k dimensions}

\noindent
Here, we construct a family of solutions to the generalized SDYM equations 
(\ref{gsdym}) which generalizes the one obtained in~\cite{IL05}. 
In the notation of section~3 we choose $\ell= k{-}1=:n$ and $r=1$.
In other words, we consider $u(2)$-valued gauge fields acting on the Fock space
$\C^2\otimes\Hcal$ with $\Hcal$ defined in  (\ref{4.8}). 
For the ADHM ingredients $\ba$, $\bb$ and $\Psi$ 
(see (\ref{Delta}) and (\ref{Psi})) we propose the ansatz
\begin{equation} \label{5.1}
{\ba}\ =\ \begin{pmatrix}
\La_1{\unity}_2 & \ldots & \La_n{\unity}_2 \\
{\zero}_2       &        & {\zero}_2       \\
                & \ddots &                 \\
{\zero}_2       &        & {\zero}_2       \end{pmatrix}
\ ,\qquad
{\bb}_\hi\ =\ \begin{pmatrix}
{\zero}_2         & \ldots & {\zero}_2         \\
b^1_\hi{\unity}_2 &        & {\zero}_2         \\
                  & \ddots &                   \\
{\zero}_2         &        & b^n_\hi{\unity}_2 \end{pmatrix}
\qquad\textrm{and}\qquad
\Psi\ =\ \begin{pmatrix}
\Psi_0 \\ \Psi_1 \\ \vdots \\ \Psi_n \end{pmatrix}\ ,
\end{equation}
where $\La_i$ and $b^i_\hj$ are real constants. Moreover, we choose
\begin{equation} \label{5.2}
b_0^i=1 \quad\textrm{for}\quad i=1,\ldots,n 
\qquad\textrm{and}\qquad b_1^1=\ldots=b^n_n=-1
\qquad\textrm{but}\qquad b^{i}_{\hat\jmath}=0 \quad\textrm{otherwise}\ .
\end{equation}
With this selection we obtain
\begin{equation} \label{5.3}
{\D}\ =\ \begin{pmatrix}
\La_1{\unity}_2 & \ldots & \La_n{\unity}_2 \\
\tilde x_1      &        & {\zero}_2       \\
                & \ddots &                 \\
{\zero}_2       &        & \tilde x_n      \end{pmatrix}
\qquad\textrm{and}\qquad
{\D^\+}\ =\ \begin{pmatrix}
\La_1{\unity}_2 & \tilde x_1^\+ &        & {\zero}_2     \\
   \vdots       &               & \ddots &               \\
\La_n{\unity}_2 & {\zero}_2     &        & \tilde x_n^\+ \end{pmatrix}\ ,
\end{equation}
where
\begin{equation}\label{5.4}
\tilde x_i\ :=\ x_0 - x_i \qquad\textrm{for}\quad i=1,\ldots,n\ .
\end{equation}

Let us plug the above ansatz into the ADHM conditions 
(\ref{c1})--(\ref{complete}).
To this end we first observe that, using (\ref{4.5}) and (\ref{4.6}), 
\begin{equation}\label{5.5}
\tilde x_i^\+ \tilde x_i \= \begin{pmatrix}
\tilde r^2_i & 0 \\[4pt] 0 & \tilde r^2_i \end{pmatrix} \und
\tilde x_i\,\tilde x_i^\+\= \begin{pmatrix}
\tilde r^2_i-2(\th_0{+}\th_i) & 0 \\[4pt] 0 & \tilde r^2_i+2(\th_0{+}\th_i) 
\end{pmatrix}\ ,
\end{equation}
where 
\begin{equation}\label{5.6}
\tilde r^2_i\ :=\ 
(\bar y_0{-}\bar y_i)(y_0{-}y_i)\ +\ (\bar z_0{-}\bar z_i)(z_0{-}z_i)\ +\
2(\th_0{+}\th_i) \= (\tilde r^2_i)^\+\ .
\end{equation}
It follows that $\ [\tilde x_i,\tilde x_i^\+]=-2(\th_0{+}\th_i)\,\s_3$.
The special case $\th_0=-\th_i\ge 0$ for $i=1,\ldots,n$ 
was considered in~\cite{IL05}. 
Here, we choose $\th_0,\th_i\ge0$ and describe more general solutions.
The operator $\tilde r^2_i$ is invertible on the Fock space
$\Hcal=\Hcal_0\otimes (\mathop{\otimes}\limits^n_{i=1}\Hcal_i)$
unless $\th_0=\th_i=0$ (the commutative case).

One can easily check that the matrices in (\ref{5.3}) satisfy
the conditions (\ref{c1}) and (\ref{c2}). Next, the equation (\ref{c3}) becomes
\begin{equation}\label{5.7}
\La_i\Psi_0 + \tilde x_i^\+\Psi_i \= \zero_2 
\qquad\textrm{for}\quad i=1,\dots,n\ ,
\end{equation}
which is solved by the `seed solution'
\begin{equation} \label{5.8}
\tilde\Psi_0\ =\ \vp_n^{-\frac12}\,{\unity}_2  \und
\tilde\Psi_i\ =\ -\tilde x_i\,\frac{\La_i}{\tilde r_i^2}\,\vp_n^{-\frac12}\ ,
\end{equation}
with $\vp_n$ to be determined. Further, the normalization (\ref{c4}) fixes 
\begin{equation}\label{5.9}
\vp_n \= 1\ +\ \sum_{i=1}^n\frac{\La_i^2}{\tilde r_i^2} \ .
\end{equation}
Hence, this seed solution $\tilde\Psi$ based on the form~(\ref{5.3}) 
for $\D$ satisfies all ADHM conditions except possibly for (\ref{complete}).

A violation of the completeness relation (\ref{complete}) simply means that 
our seed solution does not catch the full kernel of the operator $\D^\+$. To 
find the zero eigenvalues of $\D^\+$, we look for the kernel of $\tilde x_i^\+$,
\begin{equation}\label{5.10}
\tilde x_i^\+\,|B_i\> \= -\ic \begin{pmatrix}
z_0{-}z_i & \phantom{-}\bar y_0{-}\bar y_i \\[4pt] 
y_0{-}y_i & -\bar z_0{+}\bar z_i \end{pmatrix} 
\begin{pmatrix} |\b_i\> \\[4pt] |\b'_i\> \end{pmatrix} \= 0
\qquad(\textrm{no sum over}\ i) \ .
\end{equation}
For our choice $\th_0+\th_i\ge 0$,
the creation/annihilation character in (\ref{4.6}) implicates that 
the solution is given by coherent states for the corresponding 
Heisenberg-Weyl group,\footnote{
In \cite{IL05}, the special choice $\th_0=-\th_i\ge 0$ implied 
$\tilde x_i^\+\tilde x_i=\tilde x_i\,\tilde x_i^\+=\tilde r^2_i{\unity}_2$
and therefore $|B_i\>=0$, i.e. no zero modes.}
\begin{equation}\label{5.14}
|\b_i\>\= N_i\,
\exp\Bigl(\sfrac{\b_i^y(\bar y_0-\bar y_i)}{\sqrt{2(\th_0+\th_i)}}\Bigr)
\exp\Bigl(\sfrac{\b_i^z(\bar z_0-\bar z_i)}{\sqrt{2(\th_0+\th_i)}}\Bigr)\,
|0,0;*,*;\ldots;0,0;*,*;\ldots\>  
\und |\b'_i\> = 0 \ ,
\end{equation}
where $*$ indicates an arbitrary entry 
and $N_i$ normalizes to $\<\b_i|\b_i\>=1$.
Clearly, the kernel has infinite dimension. In the simplest case ($*=0$),
the Moyal-Weyl image of the projector $|\b_i\>\<\b_i|$ is a gaussian centered at
\begin{equation}\label{5.15}
(\b_i^y, \b_i^z, 0, 0,\ldots, -\b_i^y, -\b_i^z,\ldots, 0, 0)
\ \in\ \C^{2n+2}\cong\R^{4n+4}\ .
\end{equation}
For the full projector onto the kernel of $\tilde x_i^\+$ we may write
\begin{equation}\label{proj}
P_i\ :=\ |B_i\>\<B_i| \= \begin{pmatrix}|\b_i\>\<\b_i|&0\\[4pt]0&0\end{pmatrix}
\= {\unity}_2\ -\ \tilde x_i\,\frac{1}{\tilde r_i^2}\,\tilde x_i^\+ \ ,
\end{equation}
so that (no sums over $i$)
\begin{equation}\label{5.11}
\tilde x_i^\+\,\Phi_i \= 0 \qquad\Leftrightarrow\qquad
\Phi_i \= P_i\,Z_i \qquad\textrm{for}\quad i=1,\ldots,n\ ,
\end{equation}
where the $Z_i$ are arbitrary $2{\times}2$ matrices with operator entries. 
Therefore, the modified solution
\begin{equation}\label{5.12}
\Psi_0\ :=\ \tilde\Psi_0\,S^\+ \und
\Psi_i\ :=\ \tilde\Psi_i\,S^\+\ +\ \Phi_i
\end{equation}
also satisfies the equations (\ref{5.7}). 
Here, $S^\+$ is a not-yet-defined $2{\times}2$
operator-valued matrix which is allowed due to the linearity of (\ref{5.7}) 
and needed for a re-normalization of $\Psi$.

To accomplish the construction of a solution to the ADHM equations 
(\ref{c3})--(\ref{complete}), we choose 
\begin{equation}\label{5.16}
Z_i \= \begin{pmatrix}|\b_i\>\<i{-}1|&0\\[4pt]0&1\end{pmatrix}
\qquad\Rightarrow\qquad
\Phi_i \= \begin{pmatrix}|\b_i\>\<i{-}1|&0\\[4pt]0&0\end{pmatrix} 
\qquad\Rightarrow\qquad
P_i \= \Phi_i^{\phantom{|}}\,\Phi_i^\+ 
\end{equation}
with
\begin{equation}\label{5.17}
|i{-}1\>\ :=\ |i{-}1,0;0,0;\ldots;0,0\>\ \in \Hcal
\qquad\textrm{for}\quad i=1,\ldots ,n\ .
\end{equation}
Inserting (\ref{5.12}) with (\ref{5.16}) into the normalization (\ref{c4}) 
we learn that $\ S\,S^\++\sum_i\Phi_i^\+\Phi_i^{\phantom{|}}=\unity_2$, i.e.
\begin{equation}\label{5.18}
S\,S^\+\={\unity}_2\ -\ \begin{pmatrix}
\sum^n_{i=1}|i{-}1\>\<i{-}1|&0\\[4pt]0&0\end{pmatrix}
\qquad\Rightarrow\qquad
S\=\begin{pmatrix}\sum_{l\ge 0}|l{+}n\>\<l|&0\\[4pt]0&1\end{pmatrix}\ ,
\end{equation}
whence $S$ is a shift operator in the Fock space $\C^2\otimes\Hcal$.
Having defined all the `ingredients' $\tilde \Psi$, $\Phi_i$ and $S$ of the 
solution $(\D,\Psi)$, one can show by direct calculation that the completeness
relation (\ref{complete}) is now satisfied. Therefore, we can define a gauge 
potential via (\ref{adhmA}) and obtain from (\ref{adhmF}) a self-dual gauge 
field on $\R^{4n+4}_\th$.

Note that one may restrict our solution to a subspace 
$\R^{4}_{\th_\hi}\subset\R^{4n+4}_\th$. For instance, 
the potential $A_{\m_0}=\Psi^\+\pa_{\m_0}\Psi$ and its field strength
\begin{equation}\label{5.20}
F_{\m_0\n_0}\=
2\Psi^\+ \bb\,N_{\m_0\n_0}\,\bff\,\bb^\+\Psi\=
2\Psi^\+ \bb\,\begin{pmatrix}
\h_{\m_0\n_0} & \ldots & {\zero}_2 \\
   \vdots     & \ddots & \vdots   \\
  {\zero}_2   & \ldots & {\zero}_2 \end{pmatrix}
\bff\,{\bb}^\+\Psi
\end{equation}
do not contain any derivative with respect to $x^{\m_1},\ldots ,x^{\m_n}$. 
If we employ the Moyal-Weyl transform and take $\th_i\to 0$ for $i=1,\ldots,n$ 
but keep $\th_0\ne 0$, we recover the noncommutative 't~Hooft-type 
$n$-instanton solution in four dimensions as derived in~\cite{LP}. 
The role of translational moduli in this solution will be played by coordinates
$x^{\m_i}$ on the space $\R^{4n}$ which complements $\R^{4}_{\th_0}$ in 
$\R^{4}_{\th_0}\times\R^{4n}$. On the other hand, for $\th_0=-\th_i$ 
the solution (\ref{5.20}) on $\R^4_{\th_0}$ will reproduce the solution 
from~\cite{ILM}.

\bigskip

\section{Noncommutative BPST-type solution in 4k dimensions}

\noindent
In the commutative case, as it was explained in~\cite{CGK}, the 't~Hooft-type
ansatz of the previous section produces solutions of the Yang-Mills equations 
on $\R^{4k}$ which cannot be extended to the quaternionic projective space 
$\HP^k$ if $k\ge 2$. Hence, these solutions cannot have finite Pontryagin
indices $p_j$ with $j\ge 2$.  We shall now consider a different kind of ansatz
for $\ba$ and $\bb$ in (\ref{Delta}), which generates true instanton-type 
configurations (with finite Pontryagin numbers) since in the commutative limit 
$\th_{\hat\imath}\to 0$ they can be extended to $\HP^k$.

We pick the gauge group U($2k$) (i.e. $r=k$ in (\ref{Psi}) and (\ref{Delta}))
and consider the noncommutative space $\R^{4k}$ with coordinates (\ref{4.5})
satisfying the commutation relations (\ref{4.6}). Let us choose
\begin{equation}\label{6.1}
{\ba}\=\begin{pmatrix}\La\unity_2\\ \zero_2\\ \vdots\\ \zero_2\end{pmatrix}
\und
{\bb}\=\begin{pmatrix}\zero_2&\ldots&\zero_2\\ \unity_2& &\zero_2\\ 
&\ddots& \\ \zero_2&\ldots&\unity_2\end{pmatrix}
\end{equation}
as $(2k{+}2)\times 2$ and $(2k{+}2)\times 2k$ matrices, i.e. taking $\ell =1$ in
(\ref{Psi}) and (\ref{Delta}). We obtain 
\begin{equation}\label{6.2}
\Delta \= {\ba} +\bb\, \bx \=
\begin{pmatrix}\La\unity_2\\x_0\\ \vdots\\ x_{k-1}\end{pmatrix} \=
\begin{pmatrix}\La\unity_2\\[4pt] \bx \end{pmatrix}
\qquad\Rightarrow\qquad
\Delta^\+\Delta \=
\Bigl( \La^2 + \sum\limits^{k-1}_{\hi=0}x^{\m_\hi}x^{\m_\hi}\Bigr) \unity_2
\ =:\ \gamma^2\unity_2\ .
\end{equation}

Furthermore, we introduce the $2k{\times}2k$ matrix
\begin{equation}\label{6.3}
\Phi\ :=\ \unity_{2k}\ -\ \bx\,(\gamma^2{+}\La\gamma )^{-1} \bx^\+ \= \Phi^\+
\end{equation}
and the $(2k{+}2){\times}2k$ matrix
\begin{equation}\label{6.4}
\Psi \= \begin{pmatrix} -\gamma^{-1}\bx^\+ \\[4pt] \Phi \end{pmatrix}
\qquad\Rightarrow\qquad
\Psi^\+\ = (-\bx\,\gamma^{-1}\quad\Phi) \ ,
\end{equation}
which obey the relations
\begin{equation}\label{6.5}
\Phi^2 \= \unity_{2k}-\bx\,\gamma^{-2}\bx^\+\quad,\qquad 
\Phi\,\bx \= \La\, \bx\,\gamma^{-1} \und 
\bx^\+\Phi \= \La\,\gamma^{-1}\bx^\+\ .
\end{equation}
It is not difficult to see that, due to the identities (\ref{6.5}), 
the matrix $\Psi$ from (\ref{6.4}) satisfies all ADHM conditions 
(\ref{c1})--(\ref{complete}) with $\Delta$ given in (\ref{6.2}). 
Hence, for this solution $(\Delta,\Psi)$ the gauge field (\ref{adhmF}) 
satisfies the self-duality equations (\ref{gsdym}) 
(and the full Yang-Mills equations) on $\R^{4k}_\th$.
Note that for $k=1$ this solution reproduces the noncommutative one-instanton
solution as derived in~\cite{gr2}, and for $\th_0\to 0$ it coincides with the 
BPST instanton~\cite{BPST}.

In the fully commutative limit, $\th_\hi\to 0$ for $\hi=0,\ldots,k-1$,
the gauge potential $A=\Psi^\+\diff\Psi$ gives an instanton-type canonical 
$sp(k)$-connection\footnote{
Recall that $sp(k)\subset u(2k)$.} 
on the Stiefel bundle
\begin{equation}\label{st1}
\begin{CD}
\mbox{Sp}(k{+}1)/\mbox{Sp}(1) @>{\mbox{Sp}(k)}>> \HP^k
\end{CD}
\end{equation}
over $\HP^k$~\cite{NR}.
In this context, $\Psi$ of~(\ref{6.4}) is a $(2k{+}2){\times}2k$ 
matrix-valued section of this bundle. 
Clearly, in the commutative limit our solution of the generalized SDYM 
equations on $\R^{4k}_\th$ given by (\ref{adhmA}), (\ref{adhmF}) 
and (\ref{6.1})--(\ref{6.4}) turns into the instanton solution on~$\HP^k$
written in local coordinates on a patch $\R^{4k}$ of $\HP^k$.

\bigskip

\noindent
{\bf Acknowledgements}

\medskip

\noindent
The authors are grateful to A.D.~Popov for fruitful discussions and
useful comments.
T.A.I.~acknowledges the Heisenberg-Landau program and the Russian Foundation
for Basic Research (grant 06-01-00627-a) for partial support and the Institut 
f\"ur Theoretische Physik der Leibniz Universit\"at Hannover for its 
hospitality. 
The work of O.L. was partially supported by the Deutsche Forschungsgemeinschaft.
\bigskip

\end{document}